\documentstyle[manuscript,prd,aps]{revtex}
\begin{document}
\draft
\tighten
\title{CONFRONTATION OF DOUBLE-INFLATIONARY MODELS WITH OBSERVATIONS}
\author{Patrick PETER$^{1,3}$, David POLARSKI$^{2,3}$ and
A.~A.~STAROBINSKY$^{4,5}$}
\address{\hfill\\
$^1$Department of Applied Mathematics and Theoretical Physics,\\
University of Cambridge, Silver Street, Cambridge CB3 9EW (UK)\\
\hfill\\
$^2$Laboratoire de Mod\`eles de Physique Math\'ematique, EP93 CNRS\\
Universit\'e de Tours, Parc de Grandmont, F-37200 Tours (France)\\
\hfill\\
$^3$D\'epartement d'Astrophysique Relativiste et de Cosmologie,\\
Observatoire de Paris-Meudon, UPR 176 CNRS, 92195 Meudon cedex
(France)\\
\hfill\\
$^4$Yukawa Institute for Theoretical Physics, Kyoto University, Uji
611 (Japan)\\
\hfill\\
$^5$Landau Institute for Theoretical Physics, Kosygina St. 2, Moscow
117334 (Russia)\\}
\preprint{DAMTP-R94/20, YITP/U-94-6, astro-ph/9403037}
\maketitle
\begin{abstract}
We consider double-inflationary models with two noninteracting scalar
fields, a light scalar field $\phi_l$ with potential
$\frac{1}{2}m_l^2\phi_l^2$ and a heavy scalar field $\phi_h$ with
potential $\frac{\lambda}{n}\phi_h^n$ with $n=2,4$. CDM with the
initial spectrum of adiabatic perturbations produced in these models
is compared with observations. These models contain two more free
parameters than the standard CDM model with an initial scale-invariant
spectrum. We normalize our spectra to COBE DMR and compare the
predictions with observations on the biasing factor, large-scale
peculiar velocities, quasar and galaxy formation and the Stromlo-APM
counts-in-cells analysis.  The model with $n=4$ is excluded by the
data while for the $n=2$ model, taking cosmic variance into account, a
small window of parameters compatible with observations is found.
\end{abstract}

\pacs{PACS Numbers:}

\newpage
\section*{Introduction}
Inflationary models~\cite{lindekolb} can solve some of the outstanding
problems in cosmology.  Also, as emphasized already some time
ago~\cite{staro79}, for a given model it is possible to calculate the
various spectra of fluctuations produced during the inflationary phase
and to put the predictions to test using observational data. In the
simplest inflationary models, the Fourier components of the
gravitational potential are Gaussian stochastic quantities with an
approximately flat (Harrison-Zel'dovich) r.m.s.
spectrum~\cite{hawking,staro82,guth}.  For more complicated spectra,
calculation of the spectrum can be implemented either analytically or
numerically, (for the latter possibility, see for ex.~\cite{bond}),
and the spectrum can be obtained with very high accuracy.  The
(approximately) flat (Harrison-Zel'dovich) spectrum, together with the
assumptions of the standard CDM (Cold Dark Matter) model was put to
test using $N$-body simulations and a very good agreement was found
with the observed galaxy-galaxy correlation function on scales
$(0.5-10)h^{-1}$Mpc when $h\approx0.5$~\cite{davis}, see
also~\cite{sussp90} ($h\equiv H_0/100$ km/s/Mpc). Observations seem to
imply more power on scales greater than $20h^{-1}$Mpc, strong evidence
coming from the APM galaxy survey and observed peculiar
velocities~\cite{maddox} and from large-angle $\frac{\Delta T}{T}$
fluctuations~\cite{smoot92,wright,bennett}.
There are several ways to solve this
problem, either abandoning the assumptions of CDM, one may consider a
mixture of hot and cold dark matter~\cite{hcdm} or a nonvanishing
cosmological constant, or trying CDM with an initial perturbation
spectrum having more power on large scales.  Following the second
possibility, an attractive solution is to consider so called ``tilted"
spectra, with spectral index $n<1$.  Such a spectrum will occur in
power-law and in extended inflation.  Recent studies
however~\cite{cen,adams} have shown that no value of $n$ seems able to
reconcile the CDM model with all the observations (in accordance with
the earlier ``pre-COBE'' discussion in~\cite{sutherland,polarski}),
although the value
$n\approx 0.7-0.8$ comes closest to it.

So, if we want to reconcile
the CDM model with observations without introducing neutrinos with a
restmass of a few $eV$, we have to consider models belonging to the
next level of complexity, i.e. having at least one more additional
parameter characterizing the initial perturbation spectrum.
Inflationary models with flat
spectra need one free parameter to specify the amplitude of the
spectrum of density perturbations, for example the coupling constant
of the inflaton in the simplest
versions of inflation, while those generating a tilted spectrum need
one more free parameter, for example the Brans-Dicke parameter
$\omega$ in extended inflation models.  We will consider here double
inflationary
models~\cite{polarski,staro84,lev1lindestaro,lev2linde,silk,%
levpogosyan,stefanmullerstaro,zelnikov,stefanmucket}.
In the specific
models considered here, the inflationary stage is driven by scalar
fields without mutual interaction potential.
As a general rule, such models produce a spectrum of adiabatic
perturbations having more power on scales larger than some
characteristic scale~\cite{levpogosyan}. Three free parameters are now
needed: one for the height and the form of the ``step", one for its
location and finally one for the overall normalization.  We study here
two double inflationary models consisting of two noninteracting scalar
fields, a light scalar field $\phi_l$ with potential
$\frac{1}{2}m_l^2\phi_l^2$ and a heavy scalar field $\phi_h$ with
potential $\frac{\lambda}{n}\phi_h^n$. The spectra obtained
numerically are then compared with observational data. Recently, an
analogous study was performed for another double-inflationary
model~\cite{stefanmucketstaro} and we will present the results in a
way which makes the comparison easier.  In section I we give some
basic results for the models considered here and specify their free
parameters. In section II, we give the various observational tests to
which our models are submitted for different values of their
parameter. In section III finally, we give a brief discussion of our
results.
\section{The model and its fluctuation spectrum}
Let us start with a short description of the homogeneous background.
We consider the following Lagrangian density describing matter and
gravity
\begin{equation}
L=-{R\over {16\pi G}}+{1\over 2}\phi_{h,\mu}\phi_h^{,\mu}-{{\lambda}\over n}
\phi_h^n +{1\over 2}(\phi_{l,\mu}\phi_l^{,\mu}-m_l^2 \phi_l^2)
\end{equation}
where $\mu=0,..,3, c=\hbar=1$ and the Landau-Lifshitz sign conventions
are used. When $n=2,~\lambda \equiv m_h^2$, whereas for $n=4,~\lambda$
is a dimensionless coupling constant. The space-time metric has the
form
\begin{equation}
ds^2=dt^2-a^2(t)\delta_{ij}dx^i dx^j, \qquad\ i,j=1,2,3.
\end{equation}
Spatial curvature may always be neglected because it becomes
vanishingly small during the first period of inflation driven by the
heavy scalar field.  The homogeneous background is treated
classically, it is determined by the scale factor $a(t)$ and the two
scalar fields $\phi_h,\phi_l$. Their equations of motion are given by
\begin{eqnarray*}
\dot a=aH,\qquad\ H^2={{4\pi G}\over 3}\bigl ({\dot \phi_h}^2+{\dot
\phi_l}^2 +
2{{\lambda}\over n}\phi_h^n + m_l^2\phi_l^2\bigr ),\\
\ddot \phi_h+3H\dot \phi_h+\lambda \phi_h^{n-1}=0,\qquad\ \ddot \phi_l+
3H\dot\phi_l+ m_l^2\phi_l=0,
\end{eqnarray*}
where a dot denotes a derivative with respect to $t$. We have the
useful equation
\begin{equation}
\dot H=-4\pi G\bigl ({\dot \phi_h}^2+{\dot \phi_l}^2\bigr )
\end{equation}
which shows that $H$ always decreases with time in this model.
\par
Let us consider now the inhomogeneous perturbations.  We consider a
perturbed FRW background whose metric, in the longitudinal gauge,
reads
\begin{equation}
ds^2=(1+2\Phi)dt^2-a^2(t)(1-2\Psi)\delta_{ij}dx^idx^j
\end{equation}
(in Bardeen's notations~\cite{bardeen}, $\Phi=\Phi_A, \Psi=-\Phi_H$).
We get from the perturbed Einstein equations ($\exp (i{\bf {kr}})$ spatial
dependence is assumed and the Fourier transform convention is
$\Phi_k\equiv {1\over{(2\pi)^{3/2}}}\int \Phi({\bf
r})e^{-i{\bf kr}}d^{3}{\bf k}$)
\begin{eqnarray}
\Phi & = & \Psi,\nonumber\\
\dot{\Phi}+H\Phi & = & 4\pi G({\dot \phi_h}\delta\phi_h+{\dot
 \phi_l}\delta\phi_l),\nonumber\\
\delta \ddot{\phi_h} + 3H\delta \dot \phi_h + [{{k^2}\over
{a^2}}+(n-1)\lambda \phi_h^{n-2}]\delta \phi_h & = & 4\dot \phi_h
\dot \Phi -2\lambda \phi_h^{n-1} \Phi ,\nonumber\\
\delta \ddot{\phi_l} + 3H\delta \dot \phi_l + \bigl ({{k^2}\over
{a^2}}+m_l^2\bigr )\delta \phi_l & = & 4\dot \phi_l \dot \Phi
-2m_l^2\phi_l\Phi
\end{eqnarray}
We see that, contrary to the case when only one scalar
field is involved,
when we have more than one scalar field, the dynamics of the perturbed
system cannot be described by just one equation for the master
quantity $\Phi$ \cite{sasaki} or else for the gauge-invariant quantity
$\zeta=\delta
\phi+ ({\dot \phi}/H) \Phi$ in terms of which the action for the
fluctuations
can be written \cite{mukhanov}.
Analogously to~\cite{polarski} one can now compute the spectrum of
growing adiabatic perturbations.  These perturbations arise from the
vacuum fluctuations of the scalar fields $\phi_h$ and $\phi_l$. The
fluctuations are Gaussian and the power spectrum $\Phi^2(k)$ of the
gravitational potential, defined through $\langle \Phi_k
\Phi^*_{k'}\rangle=
\Phi^2(k)\delta ({\bf k}-{\bf k'})$ characterizes them completely.
For scales crossing the Hubble radius when both scalar fields are
in the slow rolling regime, the spectrum of growing adiabatic
perturbations, when those scales are outside the Hubble radius during the
matter-dominated stage (assuming $a(t)\propto t^{\frac{2}{3}}$ at the
present time), is given by
\begin{eqnarray}
k^{3\over 2}\Phi(k) & = & {6\over 5}\sqrt{2}\pi G H \sqrt{\bigl
({{2\phi_h}\over n}\bigr )^2+\phi_l^2}\\ & \simeq & {4\over
5}\sqrt{6\pi^3 G^3 \lambda {2\over n}\phi_h^n}~\phi_l
\nonumber\\
& = & {{\sqrt{24\pi G \lambda}}\over 5}\bigl ({{4\pi G}\over n}\bigr )
^{{2-n}\over 4}\sqrt{s}\ln^{n\over 4} {{k_f}\over k}
\qquad\ k\ll k_f\label{eq:log}\end{eqnarray}
where the r.h.s. has to be taken at $t=t_k$, the time at which a
perturbation with wavenumber $k$ comes outside the Hubble radius
during inflation, $k=a(t_k)H(t_k)$. The wavenumber $k_f$ corresponds
to the characteristic scale appearing in the spectrum and it is close
to the scale crossing the Hubble radius near the end of the first
inflation.
Hence, there exists a very broad interval of scales for which the
dynamics of inflation at the time of the first Hubble radius crossing
is determined by the field $\phi_h$ while the main contribution to
$\phi(k)$ is made by $\phi_l$.
We see from (\ref{eq:log}) that the upper part of the
spectrum, corresponding to small $k'$s or very large scales, is not
flat but has a logarithmic dependence $\propto \ln^{n\over 4}
\frac{k_f}{k}$. This gives, as we will see, a crucial difference
between the $n=2$ and $n=4$ spectra.  The quantity $s(t)$ is the
number of e-folds from time $t$ till the end of the second inflation
and it is given by
\begin{equation}
s \simeq 4\pi G \bigl ({{\phi_h^2}\over n}+{{\phi_l^2}\over 2}\bigr )
\end{equation}
In order to have a sufficiently long second inflationary stage that
will put the characteristic length scale of the spectrum on a scale in
agreement with observations, we have that $\phi_l\simeq 3 M_p$ near
the end of the first inflation so $s(t_f)\simeq 60$. A very little
change in this initial value will be enough to shift the spectrum in
$k$-space while leaving the form of the spectrum practically
unaltered.
We define the parameter $p\equiv {{\sqrt{\lambda M_p^{n-2}}}\over
{m_l}}$, $p^2$ gives the order of magnitude of the ratio of the energy
of the heavy scalar field to the energy of the light scalar field near
the end of the first inflation; it specifies the form of the spectrum,
namely its ``step" with more power on large scales, and the width of
the transition region in $k$-space.  We will now estimate $\Delta_k$,
the height of the ``step" in the spectrum between a scale which is on
the upper plateau (but of course still inside the cosmological
horizon) and a scale at the beginning of the lower plateau.  For
scales crossing the Hubble radius at the beginning of the second
inflationary stage (neglecting possible small oscillations), we have
the standard result of inflation driven by one scalar field in
slow-rolling regime
\begin{eqnarray}
k^{3\over 2}\Phi(k) & = & \frac{\sqrt{24\pi G m_l^2}}{5}s_0\\
& = & {4\over 5}\sqrt{6\pi^3 G^3} m_l \phi_0^2
\end{eqnarray}
where $s_0=2\pi G\phi_0^2$.
We therefore get for $\Delta_k$
\begin{eqnarray}
\Delta_k & \approx & \frac{\sqrt{\lambda}}{m_l}
\frac{\sqrt{\frac{2}{n}\phi_h^n(t_k)}}{\phi_0}\\
& \approx & \frac{\sqrt{\lambda}}{m_l}\Bigl (
\frac{4\pi G}{n}\Bigr )^{\frac{2-n}{4}}
\frac{\ln^{\frac{n}{4}} \frac{k_f}{k}}{\sqrt{s_0}} \qquad\ k\ll k_f.
\end{eqnarray}
For $n=2$, this is
\begin{eqnarray}
\Delta_k& \simeq &0.13 p \ln^{\frac{1}{2}}\frac{k_f}{k} \qquad\ k\ll
k_f\\
& \simeq &0.33 p \tilde{\phi_h}(t_k),
\end{eqnarray}
whereas $n=4$ yields
\begin{eqnarray}
\Delta_k& \simeq &0.073 p \ln{\frac{k_f}{k}} \qquad\ k\ll k_f\\
& \simeq &0.24 p \tilde{\phi_h}^2(t_k).
\end{eqnarray}
where $\tilde{\phi_h}\equiv \frac{\phi_h}{M_p}$. We will investigate
models with $3\leq p \leq 16$ for $n=4$ and $6\leq p \leq 28$ for
$n=2$.  How we choose the scales of our spectra is very important when
we compare them with the observations.  For this purpose, we need a
precise definition and we adopt here the convention
of~\cite{stefanmucketstaro}, and define it with the help of $k_b$, the
scale where the extrapolated upper part intersects the lower plateau.
One shows numerically that the evolution of the background also during
the transition between the two main inflationary stages is
inflationary, in the sense that $\ddot{a}>0$, for $p<25$ when
$n=2$~\cite{polarski2} and for $p<50$ when $n=4$. Also, the initial
perturbation spectrum will have no oscillations for $p<15$ when $n=2$
and also for all $p$'s considered here when $n=4$.  A last comment
concerns the power spectrum $P(k)$ defined through $\langle \delta_k
\delta^*_{k'}\rangle=P(k)\delta ({\bf k}-{\bf k'})$.  Linear
perturbations grow at different rates depending on the relation
between their wavelengths, the Jeans length and the Hubble radius and
this is specified by the transfer function $T(k)$ [more accurately,
one should write $T(k,t_0)$]
\begin{equation}P(k,t_0)=\frac{4}{9}\frac{k^4}{H_0^4}\Phi^2(k)T^2(k).
\label{eq:transfer}\end{equation}
where by definition, $T(k\rightarrow 0)=1$. $T(k)$ is computed numerically
making assumptions about the matter content of the universe, and
depends on
parameters like $\Omega_0$ and $h$. We use
here the transfer function for the standard CDM model given by~\cite{bbks}
\begin{equation}T(q)=\frac{\ln (1+2.34q)}{2.34q}\lbrack 1+3.89q+(16.1q)^2+
(5.46q)^3+(6.72q)^4\rbrack^{-\frac{1}{4}}\end{equation} where $q\equiv
\frac{k}{\Omega_0 h^2 Mpc^{-1}}$, $\Omega_0=1$, $h=0.5$. We will
assume tacitly in all the formulas that the density parameter
$\Omega_0=1$, in accordance with inflation. The power spectra $P(k)$ for
different values of the parameters are displayed for the $n=2$, resp.
$n=4$ model in Fig.1, resp. Fig.2.

\section{Confrontation with observations}

\subsection{Normalization of the spectrum to COBE DMR}\label{sub1}

Let us start with the normalization of the spectrum. Since the
discovery of large angular scale fluctuations in the Cosmic Microwave
Background Radiation by COBE DMR~\cite{smoot92}, one can normalize the
spectrum of fluctuations using these COBE DMR observations.
These observations probe the spectrum of fluctuations on very large
scales, from several hundreds Mpc up to the cosmological horizon and
allow for a normalization based on first principles.  More precisely,
one has
\begin{equation}\sigma_{T}^2(10^0)= \sum_{l\geq 2}\frac{2l+1}{4\pi}
\langle |a_{lm}|^2 \rangle \exp [-l(l+1)\theta_0^2]\equiv \sum_{l\geq 2}a_l^2
\exp [-l(l+1)\theta_0^2],\end{equation}
where $\theta_0=0.425 \theta_{FWHM}=4.25^0$ is the Gaussian angle
corresponding to the antenna beam and additional smearing of the raw
data.  It is to be noted that due to the exponential factor, only the
multipoles up to $l\leq 20$ contribute significantly to the observed
anisotropy.
The r.m.s. coefficients $\sqrt{\langle |a_{lm}|^2 \rangle}$, with
$a_{lm}$ defined by
\begin{equation}\frac{\Delta T}{T}=\sum_{l,m}a_{lm}Y_{lm}
\end{equation}
are actually independent of $m$. Based on the latest results of COBE
DMR~\cite{wright,bennett}, we take $\sigma^2_T(10^0)=(1.25\pm 0.2)\times
10^{-5}$, with error bars at the 1$\sigma$--level. This allows us to
normalize the fluctuation
spectrum and constitutes a great step forward.  For the CMBR
anisotropy, on large angular scales the dominant effect is the
Sachs-Wolfe effect and, for adiabatic perturbations, we have the
fundamental relation
\begin{equation}\langle |a_{lm}|^2
\rangle=\frac{H^4_0}{2\pi}\int_0^{\infty}
dk k^{-2} P(k) j_l^2(kr_{rec}),\label{sw}\end{equation}
where $j_l$ is
a spherical Bessel function and $r_{rec}$ is the comoving distance
between us and the surface of recombination, we have in very good
approximation $r_{rec}=\frac{2}{H_0}$.  In the models we are
considering here, normalization of the power spectrum $P(k)$, which is
obtained for {\it given $p$ and location of the ``step" }, through eq.
(\ref{sw}) will fix the value of the remaining free parameter
$\lambda$ or equivalently $m_h$. The parameter $m_h$ will have the
order of magnitude given thereafter when we vary $p$ and $k_b$
earlier~\cite{polarski2}. As a result, the energy density at the
beginning of the second inflation is also fixed: it is $\sim
\frac{5}{p^2}\times 10^{-11}\mbox{M}^4_p$ for $n=2$. For the parameter
$\lambda$, we get the following result, when we vary the
parameters: $\sqrt{\lambda}\sim 2\times
10^{-6}$ and the corresponding energy density is then $\sim
\frac{2}{p^2}\times 10^{-11}\mbox{M}_p^4$ for $n=4$.  Another
remark concerns the contribution to the CMBR anisotropy on large
scales ($l\leq 40$) which comes from gravitational waves (tensor metric
perturbations). One can show that this contribution is equal for both
models and rather small, namely $\sqrt{\langle
|a_{lm}|^2\rangle_{tot}}
\equiv \sqrt{(1+\frac{T}{S})\langle |a_{lm}|^2\rangle_{AP}}
\approx 1.05 \sqrt{\langle |a_{lm}|^2 \rangle_{AP}}$
where the subscript $AP$ refers to that part of the fluctuations due
to adiabatic perturbations, i.e., the
quantity calculated in~(\ref{sw}). Although this effect is not large,
it is important here for the $n=2$ model, which for some choices of
the parameters will turn out to be in marginal
agreement with observations. Also, one has to take into account
``cosmic variance" which is connected to the fact that we try to
estimate r.m.s. values of physical quantities in the universe
from a limited sample. Therefore, independently on how good the COBE
measurements may be, the quantities
$\sum_{l=-m}^m\frac{|a_{lm}|^2}{\langle |a_{lm}|^2 \rangle}$ obey
a $\chi^2$-distribution for $2l+1$ d.o.f.

In connection with recent interest in relations between
$\frac{T}{S}\equiv \frac{\langle |a_{lm}|^2\rangle_{GW}}
{\langle |a_{lm}|^2\rangle_{AP}}$ (for $1\leq l\leq 40$) and the
slopes
of power spectra of adiabatic perturbations,
$n_S=1+\frac{d\log(k^3\Phi^2(k))}{d\log k}$, and of
gravitational waves, $n_T=\frac{d\log(k^3\Phi^2(k))}{d\log k}$, it is
interesting to note that the relation $n_T\approx n_S-1$ (proposed,
e.g., in~\cite{critt1,critt2}) is valid in our models for $k\ll k_f$
(and not too small) though it is not valid for $k>k_f$ and also not
valid for single chaotic inflation. On the other hand, the
relation $\frac{T}{S}\approx 6.2 n_T$ which is valid for single
slow-rolling inflation (see e.g.~\cite{staroGW} for chaotic inflation)
is strongly violated if $k_{hor}\ll k_f$ (this issue will be addressed
in a separate publication~\cite{polarski3}).
%
%
\subsection{Large-scale peculiar velocities}\label{sub2}

A lot of information can be gained from the observation of peculiar
velocities, velocities in addition to the Hubble flow. As all matter
contributes gravitationally, the peculiar velocities sample all the
mass and not just the galaxies. Hence, knowing the peculiar velocity
field would give us an information on the primordial spectrum of the
same interest as the CMBR anisotropy. It should however be stressed
that large-scale peculiar velocities have rather large uncertainties.
In the linear regime, gravitational instability produces a velocity
field that is irrotational at sufficiently late times, the velocity
field ${\bf v}$ then derives from a velocity potential $\Psi$ with
${\bf v}=-\vec{\nabla}\Psi$. Measurements of redshifts $z$ and of
galaxy distances $r$ (actually of $H_0r$) provide the radial component
of the peculiar velocity field:$v_r=cz-H_0r$, for a galaxy at small
redshift.  For a potential flow, the radial velocity field when
integrated along radial paths gives the velocity potential out of
which the other velocity components can be derived. The difficulty
with this method is to construct a smooth radial velocity field and to
eliminate the statistical uncertainties in $v_r$. This is done with
interpolation and smoothing of the raw data and we finally have the
following equation
\begin{equation}\langle v^2
\rangle_R=\frac{H^2_0}{2\pi^2}\int_0^{\infty}dk
P(k) W^2_{TH}(kr)\exp(-kR_s)\end{equation}
where $W_{TH}(kR)$ stands for the Fourier transform, up to a constant,
of the ``Top Hat" window function
\begin{equation}W(kR)=\frac{3}{(kR)^3}\bigl (\sin kR-kR\cos kR\bigr ).
\end{equation}
We will compare our predictions with the Potent data. In these
observations the raw data are smoothed with a Gaussian smoothing
radius $R_s=12h^{-1}$Mpc while spheres of radii $R=40h^{-1}$Mpc and
$R=60h^{-1}$Mpc were considered. The data are $v_{40}=405\pm 60$ km/s
and $v_{60}=340\pm 50$ km/s~\cite{dekel}, error bars at the 1$\sigma$
level.  Analogously to what was found in~\cite{stefanmucketstaro},
this test is crucial for our models too.

Velocities are generally too low and all the models are excluded if
one doesn't invoke cosmic variance for the measured peculiar
velocities. For the $n=2$ models, the best velocities are
systematically obtained for $\frac{2\pi}{k_b}\sim 6h^{-1}$Mpc. This is
also the case for $n=4$, except when $p<6$, but the velocities do not
grow significantly for smaller $k_b$. Also, it turns out that the
$n=4$ model can be excluded: for $p<5$ one gets the best velocities,
but other tests exclude these models while for $p\geq 5$, the
velocities become too small.

\subsection{The biasing factor b}\label{sub3}

Before the COBE DMR observations of the CMBR anisotropy, one way to
normalize the fluctuation spectrum was through the quantity
$\sigma_8\equiv \langle \bigl (\frac{\delta M}{M}\bigr
)^2\rangle_{R=8h^{-1}Mpc} $. This quantity measures the variance of
the total mass fluctuations in a sphere of radius $R=8h^{-1}$Mpc. The
reason for considering spheres of radii $R=8h^{-1}$Mpc is that for
bright galaxies $\sigma_8$ is equal to one \cite{peebles}. However,
one doesn't expect the total matter to be as clustered as bright
galaxies and one tries therefore the simplest assumption, namely that
there is a scale independent bias, given by the biasing factor $b$
\begin{equation}\sigma^2_{R,g}=b^2\sigma^2_R \qquad\ \xi_{gg}=b^2\xi
\end{equation}
where the subscript $g$ refers to galaxies while $\xi$ is the two-point
correlation function.
$\sigma^2_R$ can be computed from the power spectrum $P(k)$
\begin{equation}\sigma^2_R=\frac{1}{2\pi^2}\int_0^{\infty}dk k^2
W^2_{TH}(kR)
P(k).\label{eq33}\end{equation}
Early numerical simulations of CDM models ~\cite{davis} were
able to reproduce the correct two-point correlation function $\xi$
which is certainly a great success for CDM. It required however, if one
imposes $\Omega_0=1$, that $h$ is as low as $h\sim 0.25$ . However for
$h\approx 0.5$ one gets the right amplitude for $\xi$ assuming that
$b\approx 2$. Another reason for considering a bias comes from the
observed r.m.s. peculiar velocities between galaxy pairs.  If
$\Omega_0=1$ and there is no bias then numerical simulations give
velocities $\sim 1000$km/s, as expected also on theoretical grounds,
much larger than the observed ones $\sim 300\pm 50$km/s (see
however~\cite{couchman} for a possible velocity bias). Here also, an
$\Omega_0 =1$ model will be compatible with the observed peculiar
velocity fields if we introduce a biasing parameter $b\approx 2.5$.
Finally, $b=2.0\pm 0.2$ is needed in order to get the correct amount
of clusters of galaxies~\cite{bahcall}.

For $n=4$, we get an unacceptably low $b$ ($b<1.5$) for $p<5$, for
$p=5,~b\approx 1.5$. For $n=2$, we get too low $b$ for $p<8$ which are
just the values that produce the best velocities.

\subsection{Formation of galaxies and quasars}\label{sub4}

The very existence of compact objects observed at high redshifts
constrains any proposed model for the formation of galaxies as these
objects must already have formed at these high $z$. Quasars have now
been observed up to redshifts near $z=5$. They are believed to be
powered by massive black holes located at the center of galaxies. From
luminosity bounds one can estimate the mass of the black holes
($M\approx 10^9M_{\odot}$), while for the host galaxy the estimates give
$M\approx 10^{11}-10^{12}M_{\odot}$. Although the formation of
gravitationally bound objects is a complicated non-linear process one
can, using rather simple assumptions, make a connection with the
linear theory \cite{sussp90,liddle93,padma93}. The fraction $F(>M)$ of
bound objects with mass greater than some given mass $M$ can be
expressed as a function of $\sigma_R$~\cite{press74}, where $R$ is the
radius of the sphere containing an amount of mass $M$ today, and
$\sigma_R$ is calculated assuming a linear evolution (in this
subsection we will adopt the notation $\sigma(M)$:
\begin{equation}F(>M)= \mbox{erfc}(\frac{\nu}{{\sqrt{2}}})
\label{eq:press}\end{equation}
where erfc is the complementary error function, $\nu=\delta_c(1+z)
\sigma^{-1}(M)$. The spherical collapse model gives for the collapse
threshold
value $\delta_c=1.686$, the value used here, though estimates from
numerical simulations suggest other acceptable values ($1.33\leq
\delta_c \leq 2$).  The most recent estimates of the mass fraction
$F(10^{11}M_{\odot})$ in host galaxies of quasars at
$z=4$~\cite{haehnelt} yields the
following lower bound~\cite{stefanmucketstaro}, using~(\ref{eq:press})~:
\begin{equation}\sigma (10^{11}M_{\odot})\approx 2.2\pm 0.5\label{quasar}
\end{equation}
When comparing our predictions with the data, we should keep in mind
that equation~(\ref{quasar}) is only a lower bound as quasars do not
necessarily form in all potential host galaxies and the real number of
quasars at $z=4$ might be larger than the observed one.  Also
important is the fact that many large galaxies seem to have formed
already at $z=1$. From this observation we get the lower bound
\begin{equation}\sigma (10^{12}M_{\odot})\approx 2.0\pm 0.4\label{gal}
\end{equation}
Approximately the same estimate follows from Ref.~\cite{haehnelt}

For $n=2,~p\geq 25$, considering the best velocities obtained for
$\frac{2\pi}{k_b}\approx 6h^{-1}$Mpc, the values obtained for this
test are too low. For these models, it is interesting to point out
that if the overall normalization goes up, hence improving these
numbers, the biasing factor of these models will become too low so
that these models must be rejected.

\subsection{Counts-in-cells analysis}\label{sub5}

We finally compare our models with the counts-in-cells analysis of
large- scale clustering of the Stromlo-APM redshift
survey~\cite{loveday92}. Values for the counts-in-cells variance
$\sigma^2_l$, where $l$ is the cell size expressed in $h^{-1}$Mpc,
obtained with our spectra normalized according to $\sigma_8=1$,
corresponding to optical galaxies in redshift space, are compared with
the Stromlo-APM data.  In order to decide whether our model fits the
data well, we apply a $\chi^2$ analysis. Considering the 9 data points
(for 9 different cell sizes) as independent and the error bars quoted
in \cite{loveday92} as $2\sigma$ ones, while we test here a theory
with 2 parameters $p$ and $k_b$ (we still have the possibility to
change the normalization, this is just changing $b$ ), we have a
$\chi^2$ distribution with 7 d.o.f. The variance $\sigma^2_l$ can be
written as
\begin{equation}\sigma^2_l=\frac{1}{2\pi^2}\int_0^{\infty}dk k^2 W^2_c(kl)
P(k).\label{eq34}\end{equation}
where $W^2_c$ is the analogous of $W^2_{TH}$, but now for a cell of size $l$
\begin{equation}W^2_c(kl)=8\int_0^1dx\int_0^1dy\int_0^1dz(1-x)(1-y)(1-z)
\frac{\sin(klr)}{klr} \qquad\ r\equiv |\vec{x}|.\end{equation}
A $\chi^2<7$ will be considered good while $\chi^2>18$ will be
considered bad.

The $n=4$ model gives very bad numbers, $\chi^2>30$ for $p\leq
5,~\chi^2>20$ for $p< 8$, for $p=8$ the test is still not too good,
$\chi^2>11$. All the models with $n=2,~p>10$ will yield very good
results, $2<\chi^2<3$ for $\frac{2\pi}{k_b}\approx (6-10)h^{-1}$Mpc.

\section{Discussion and conclusion}

The obtained large scale peculiar velocities are low for all values of
$p$ and $k_b$. A possible solution to this problem is to assume that,
due to ``cosmic variance", the velocities observed around us are
higher than the real average. If we assume that the peculiar
velocities are measured from just one independent volume, $v^2$
(actually $v^2_{40}$ or $v^2_{60}$) itself obeys a
$\chi^2$-distribution (with one d.o.f. and variance
$\sigma^2(v^2)=2$), for example there is a 20\% probability to have
$v^2\geq 1.6\langle v^2 \rangle$ and a 30\% probability to have
$v^2\geq 1.1\langle v^2 \rangle$. But one has to be cautious, when
invoking cosmic variance since a $\chi^2$-distibution with one d.o.f.
for $v^2$ might be too crude.

When $n=4$, we get the highest, though still rather low, velocities
for the lowest values of $p$. But for $p<5$, we get unacceptably low
$b$ and unacceptably high values for $\sigma(10^{11}M_\odot)$ and
$\sigma(10^{12}M_\odot)$. Comparison with the counts-in-cells analysis
gives very bad results for $p<8$.
For $p=6,8$ we get acceptable
$\sigma(10^{11}M_\odot)$ and $\sigma(10^{12}M_\odot)$, however the
velocities become unacceptably low and this situation becomes only
worse with growing $p$. For example $p=14$ gives unacceptably high $b$
and unacceptably low $\sigma(10^{11}M_\odot),~\sigma(10^{12}M_\odot)$
and $v$.  Hence, the $n=4$ model is very unlikely and can be
confidently excluded.

An improved situation is obtained when $n=2$. We obtain a window of
allowed parameters for $\frac{2\pi}{k_b}\sim (6-10)h^{-1}$Mpc and
$10<p<15$ (see fig.3). However, also for this window, the velocities
are too low
compared to the POTENT data and one has to invoke ``cosmic variance"
and higher observed velocities around us than the real average one.
In order to avoid that the magnitude of this effect is too unprobably
high,
we can also push the COBE data to their upper $(1-1.5)\sigma$ error
bar, an increase of about (20-25)\% resulting in the same increase in
the power spectrum $P(k)$ and a corresponding decrease of the biasing
factor $b$ ($\chi^2$ is unaffected). We would still get acceptable
numbers, though $b$ would become rather low ($\approx 1.5$), in
particular $\sigma(10^{11}M_\odot),~\sigma(10^{12}M_\odot)$ which are
otherwise a little bit low get better.
Note however, that even for the mean COBE normalization, the values of
the large-scale bulk velocities in the $n=2$ model for the allowed
range of parameters are slightly higher than those in the CDM model
with a cosmological constant and flat initial perturbation spectrum
(see e.g.~\cite{kgb}) which are in turn higher than those in the
tilted CDM model with $n_S\leq 0.9$. Thus, this problem is less
severe for the double inflationary $n=2$ model than for the latter
models.

In conclusion, we have considered here two double-inflationary models
with two noninteracting scalar fields: a light scalar field $\phi_l$
with potential $\frac{1}{2}m_l^2\phi_l^2$ and a heavy scalar field
$\phi_h$ with potential $\frac{\lambda}{n}\phi_h^n$. We have analized
numerically the cases $n=2$ and $n=4$. Trying CDM with an initial
spectrum produced by these models, one has three free parameters, that
is two more than the standard CDM model with a (approximately)
scale-invariant initial spectrum.  For given $k_b$ (location of the
``step") and given $p\equiv \frac{\sqrt{\lambda M_p^{n-2}}}{m_l}$,
normalization to COBE DMR constrain the remaining free parameter of
the model: for $n=2, ~m_h\sim3\times 10^{-6}$M$_p$, a result already
reported earlier~\cite{polarski2}, for $n=4,~\sqrt{\lambda}\sim
2\times 10^{-6}$. In this way also, the energy scale of the
inflationary phase is determined, scales $\sim \frac{2\pi}{k_b}$ cross
the horizon (for the first time) close to the beginning of the second
inflation corresponding to an energy density $\sim \frac{5}{p^2}\times
10^{-11}\mbox{M}_p^4$ for $n=2$ and $\sim \frac{2}{p^2}\times
10^{-11}\mbox{M}_p^4$ for $n=4$. As was recently emphasized in the
study of
another double-inflationary model~\cite{stefanmucketstaro},
here too the peculiar velocities
obtained are very small for all values of the parameters and this
poses a severe constraint on the model. The $n=4$ model is shown to be
excluded while the $n=2$ model is marginally admissible for the range of
parameters $\frac{2\pi}{k_b}\sim (6-10)h^{-1}$Mpc and $10<p<15$. In
the latter case, the remaining difficulty is still with low
large-scale bulk velocities though it is less severe than in the
CDM+$\Lambda$ model or the tilted CDM model.


If one assumes, invoking cosmic variance,
that the average values of the measured peculiar velocities are higher
than their actual r.m.s. values, and taking the upper $(1-1.5)\sigma$
limit of the COBE DMR data for the overall normalization, a window of
parameters mentioned above for the $n=2$ model remains compatible with
observations, however with
velocities about 30\% less than the lower 1$\sigma$ measured ones. If
the COBE DMR
measurements go a little bit up as they did recently, while the
measured peculiar
velocities on which there are still rather large uncertainties godown,
then these models will do better. Another potential difficulty which
might also be cured by an overall increase of amplitude is the rather
small total density fluctuations at galaxy scales.

\vspace{20pt}
\par
\section*{Acknowledgments}
D.P. would like to thank L. Kofman for stimulating conversations.
A.S. and D.P. are grateful to Profs Y. Nagaoka and J. Yokoyama for their
hospitality at the Yukawa Institute for Theoretical Physics, Kyoto
University. The financial support for research work of A.S. in Russia
was provided by the Russian Foundation for Basic Research, Project
Code 93-02-3631, and by the Russian Research Project
``Cosmomicrophysics''. P.P is supported by SERC grant \# 15091-AOZ-L9.

\begin{figure}
\caption{Power spectrum for $n=2$ and, from top to bottom on the left,
  $p=$12, 10, 15 and 28 (full lines), compared to a scale invariant
spectrum (dashed line) while all spectra are normalized to COBE.}
\end{figure}

\begin{figure}
\caption{Power spectrum for $n=4$ and, from top to bottom on the left,
$p=$3, 8 and 16 (full line), again with a scale invariant spectrum
(dashed) and all spectra normalized to COBE.}
\end{figure}

\begin{figure}
\caption{Schematic representation of the behaviour of the $n=2$ model
in the $p-k_b$ plane with normalization according to the mean COBE
data. For $p<8,~b<1.5$ while for $p>14,~\sigma(10^{12}M_\odot)<1.5$.
Above the dashed line, $v_{60}<215$km/s. Inside the upper, resp. lower curve,
$\chi^2<7$, resp. 3.}

\end{figure}

\begin{table}
\begin{center}
\begin{tabular}{|l|d|d|d|d|d|d|}
$p$ & 8 & 10 & 12 & 14 & 15 & 28 \\
\hline
$ b $ & 1.53 & 1.78 & 1.85 & 1.94 &1.89 &1.63 \\
\hline
$ v_{40}$ & 269 & 260 & 263 & 263 & 266 & 282 \\
\hline
$ v_{60}$ & 221 & 215 & 218 & 218  & 220 & 231 \\
\hline
$ \sigma (10^{11} M_\odot )$ & 3.64 & 2.85 & 2.32 & 1.99 & 1.88 & 1.4
\\
\hline
$ \sigma (10^{12} M_\odot )$ & 2.69 & 2.1 & 1.75 & 1.50  & 1.44 & 1.26 \\
\hline
$\chi^2$ & 17.23 & 8.98 & 4.35 & 2.08 & 2.13 & 4.47 \\
\hline
\end{tabular}
\end{center}
\caption{Values of the various tests for the model with $n=2$. All
these number have been calculated with $k_b^{-1}=1 h^{-1}$~Mpc.}
\end{table}

\begin{table}
\begin{center}
\begin{tabular}{|l|d|d|d|d|d|}
$p$ & 3 & 5 & 6 & 8 & 16 \\
\hline
$ b $ &1.16 & 1.56 & 1.78 & 2.2 & 3.09 \\
\hline
$ v_{40}$ & 298 & 250 & 235 & 217 & 208 \\
\hline
$ v_{60}$ & 241 & 206 & 195 & 182 & 176 \\
\hline
$ \sigma (10^{11} M_\odot )$ & 5.49 & 3.83 & 3.22 & 2.38 & 1.11 \\
\hline
$ \sigma (10^{12} M_\odot )$ & 4. & 2.82 & 2.38 & 1.78 & 0.87 \\
\hline
$\chi^2$ & 41.88 & 27.39 & 23.24 & 11.39 & 3.85 \\
\hline
\end{tabular}
\end{center}
\caption{Values of the various tests for the model with $n=4$. All
these number have been calculated with $k_b^{-1}=1 h^{-1}$~Mpc.}
\end{table}

\begin{table}
\begin{center}
\begin{tabular}{|l|d|d|d|d|}
$k_b$ & 1 & 1.2 & 1.5 & 3\\
\hline
$ b $ & 1.97 & 2.07 & 2.21 & 2.64 \\
\hline
$ v_{40}$ & 260 & 255 & 250 & 228 \\
\hline
$ v_{60}$& 215 & 212 & 208 & 193 \\
\hline
$ \sigma (10^{11} M_\odot )$ & 2.14 & 2.15 & 2.16 & 2.25 \\
\hline
$ \sigma (10^{12} M_\odot )$ & 1.6 & 1.58 & 1.58 & 1.68 \\
\hline
$\chi^2$ & 2.51 & 1.85 & 1.97 & 3.4 \\
\hline
\end{tabular}
\end{center}
\caption{$n=2$ model for $p=13$ and values of $k_b^{-1}$, expressed in
$h^{-1}$~Mpc, around the window of best parameters.}
\end{table}

\begin{table}
\begin{center}
\begin{tabular}{|l|d|d|d|d|d|d|d|d|d|}
$l$ & 10 & 15 & 20 & 25 & 30 & 40 & 50 & 60 & 75 \\
\hline
$\sigma^2(l)$ & 1.346 & 0.773 & 0.488 & 0.327 & 0.230 & 0.126 & .0738
& .0486 & .0272 \\
\hline
$\sigma^2_{obs}(l)$ & 1.24 & 0.74 & 0.49 & 0.37 & 0.24 & 0.14 & 0.080
& 0.048 & 0.025 \\
\hline
\end{tabular}
\end{center}
\caption{Comparison between predicted $\sigma^2(l)$, where $l$ is the
cell size expressed in $h^{-1}$Mpc and $\sigma^2_{obs}(l)$, the values
inferred from the Stromlo-APM redshift survey, for $n=2$, $p=13$ and
$k_b^{-1}=1 h^{-1}$~Mpc.  For the values displayed, the $\chi ^2$ test
gives $\chi ^2 = 2.51$.}
\end{table}

\end{document}